\documentclass[final,5p,times,twocolumn]{elsarticle}
\journal{Physics Letters B}

\usepackage{graphicx} 
\usepackage{subfigure}
\usepackage{amssymb}
\usepackage{units}
\usepackage{xcolor}
\usepackage{amsmath}
\usepackage{units}
\biboptions{compress}
\usepackage{hyperref}

\newcommand{\Fig}[1]{{Fig.~\ref{#1}}}

\begin{document}

\begin{frontmatter}

\title{Colour coherence in small collision systems}

\author[add1,add2,add4]{Isobel Kolbé}
\author[add3]{Chiara Le Roux}
\author[add3]{Korinna Zapp}
\affiliation[add1]{organization={School of Physics, University of the Witwatersrand},
            addressline={1 Jan Smuts Ave}, 
            city={Braamfontein},
            postcode={2000},
            country={South Africa}}
\affiliation[add2]{organization={Mandelstam Institute for Theoretical Physics, University of the Witwatersrand},
            addressline={1 Jan Smuts Ave}, 
            city={Braamfontein},
            postcode={2000},
            country={South Africa}}
\affiliation[add4]{organization={National Institute for Theoretical and Computational Sciences},
            addressline={Merensky building, Merriman Street}, 
            city={Stellenbosch},
            postcode={7600},
            country={South Africa}}
\affiliation[add3]{organization={Dept. of Physics, Lund University},
            addressline={Sölvegatan 14A}, 
            city={Lund},
            postcode={S22362}, 
            state={},
            country={Sweden}}

\begin{abstract}

The observation of collectivity in collisions of small systems has constituted a challenge for the heavy-ion community for over a decade now. The absence of jet quenching in those systems presents an apparent contradiction to the presence of an azimuthal anisotropy of high-$p_\perp$ particles. In the present work, we investigate the role of colour coherence in this puzzle. For that, we use the \textsc{Jewel} Monte Carlo model in its latest version, which includes effects of colour coherence in the jet-medium interactions. We then compare the two scenarios, with and without colour coherence, and quantify the effect on hadron and jet $R_{AA}$ as well as on high-$p_\perp$ $v_2$. The results indicate that, although coherence effects do account for an increase in $R_{AA}$, they do not affect $v_2$ to the same extent. Using hydrodynamic profiles generated with \textit{Trajectum} we compare O+O and Pb+Pb collisions at the same charged particle multiplicity and find that the nuclear modification factors are the same in both systems despite their different shapes. 

\end{abstract}


\end{frontmatter}

\section{Introduction}

The formation of a medium of quarks and gluons in heavy-ion collisions was proposed \cite{Wiedemann:2009sh,Majumder:2010qh,Mehtar-Tani:2013pia,Qin:2015srf,Apolinario:2022vzg} and observed \cite{Connors:2017ptx,Cunqueiro:2021wls,ALICE:2022wpn,Hayrapetyan2024CMSHeavyIon} for a long time now. In this process, high transverse momentum hadrons and jets have played an important role since they are modified by losing energy and momentum via successive interactions with the medium particles \cite{Apolinario:2022vzg}. The same process of path-length dependent energy loss also leads to an azimuthal anisotropy of high-$p_\perp$ hadrons and jets in the final state. Both of these signatures of a medium have been observed in the collision of heavy ions. Nonetheless, when it comes to small systems, no energy loss has been observed whereas the anisotropy of high-$p_\perp$ particles is present ~\cite{CMS:2016_pPbJets,Oh:2023_HFJets,ATLAS:2019vcm,ALICE:2022cwa,CMS-PAS-HIN-23-002}.

To reconcile these contradictory observations, it is important to understand the dynamics of partonic energy loss. One important aspect in this process is that hard partons carry colour charge and, therefore, are subject to the effects of colour coherence. For example, a parton forming a color dipole may not be resolved by the medium as an independent particle depending on the size of the dipole. Therefore, the re-scattering of this parton is suppressed relative to the incoherent case. Since this effect relies on the size of the dipole, it is expected that it is more important at early times when the dipoles are typically smaller than later on. As a result, this effect could be more important in collisions of small systems compared to large ones. The question thus arises to what extent colour coherence could be responsible for the apparent absence of jet quenching in small collision systems, and whether it affects nuclear modification factors quantifying the overall suppression and anisotropic flow coefficients in the same way.

The present work takes a first step in addressing theses questions and explores the size of colour coherence effects in O+O collisions, their sensitivity to re-scattering at early times, and the relation between nuclear modification factors and elliptic flow coefficients for high transverse momentum hadrons and jets within the \textsc{Jewel} framework for jet evolution in a dense background medium.

\section{Model}

For the work presented in this paper, we have generated events using the \textsc{Jewel} Monte Carlo model \cite{Zapp:2012ak,Zapp:2013vla}. \textsc{Jewel} is based on \textsc{Pythia}\,6.4~\cite{Sjostrand:2006za}, which is used for the hard matrix elements, initial state parton showers, and the hadronisation. \textsc{Jewel} has its own virtuality ordered final state parton shower, which is constructed with off-shell kinematics such that there is no need to correct the kinematics of an earlier splitting \textit{a posteriori}. In the presence of a medium of coloured particles, the partons can re-scatter off medium particles according to $t$-channel pQCD matrix elements for elastic scattering regularised by the Debye mass. Medium induced emissions are taken care of by resetting the parton shower at a higher scale whenever a scattering occurs at a scale above that of the parton shower and the formation time of the first emission from the new parton shower has a shorter formation time than that from the old parton shower. The Landau-Pomeranchuk-Migdal effect~\cite{Landau:1953gr,Migdal:1956tc} is also included by treating scatterings falling within the formation time of a given splitting coherently~\cite{Zapp:2008af}. \textsc{Jewel} can also keep the medium particles that scattered off the hard partons in order to account for medium response. All these particles are then hadronised using the \textsc{Pythia} string model.

For the present work, the latest version 2.6.0\footnote{The code will soon be available at \url{jewel.hepforge.org}.}~\cite{ccpaper} was used. This way, angular ordering can be imposed dynamically only for those splittings that take place sequentially with no medium interactions in between. Moreover, the code includes a new mechanism to take care of colour coherence. This follows qualitatively the findings of~\cite{Mehtar-Tani:2010ebp,Mehtar-Tani:2011hma,Casalderrey-Solana:2011ule} and involves rejecting medium scatterings off single partons that happen at scales $1/q_\perp$ smaller than the size of the colour dipole which that parton is a part of. On the other hand, the two partons in a dipole may scatter coherently with the medium. This is done by performing the scattering of the summed four-momentum of the two partons and then splitting the resulting vector again. The model was run with default settings without any tuning.

The simulations presented here were carried out with the EPPS21~\cite{Eskola:2021nhw} nuclear PDF sets on top of the CT18ANLO~\cite{Hou:2019efy} proton PDF set, which was also used for the p+p baseline calculations. Medium response is included with the subtraction of thermal momenta performed using the Constituent subtraction procedure~\cite{Milhano:2022kzx}.

\smallskip

The parton shower evolution and re-scattering is fairly agnostic to the medium model. Therefore, \textsc{Jewel} can be interface with several different media. For the present work, we have used two different models for the medium: the default simplified model that is deployed with \textsc{Jewel} supplemented by a Glauber model \cite{lisbonThesis,lisbonThesis2,fluctations} to obtain the initial temperature profile. It uses the TGlauberMC~\cite{Loizides:2017ack} implementation of the \textsc{Phobos} MC Glauber model~\cite{Alver:2008aq} to obtain the positions of wounded nucleons and binary nucleon--nucleon scatterings on an event-by-event basis. A Gaussian energy density profile is placed at the position of each wounded nucleon in the transverse plane at the initialisation time $\tau_i$, which then gets translated into a temperature profile. The medium is assumed to be boost invariant along the beam direction and the temperature decreases according to $T(x,y,\tau) = T(x,y,\tau_i)(\tau/\tau_i)^{-1/3}$ due to the longitudinal expansion. In this model, the system can expand longitudinally but no transverse expansion takes place. The model has two free parameters, namely the initialisation time $\tau_i$ and the average temperature $T_i$ at $\tau_i$ in the centre ($x=y=0$) of a central (impact parameter $b=0$) collision.

Alternatively, we have also used the hydrodynamic medium model \textit{Trajectum} \cite{Nijs:2020roc}.
\textit{Trajectum} is an open source heavy-ion simulation framework which combines models of the initial condition, pre-equilibrium evolution, viscous hydrodynamics, and hadronic transport.
\textit{Trajectum} does not contain a high-$p_T$ parton shower, but can be interfaced with \textsc{Jewel} using the publicly available interface \cite{Kolbe:2023jba}.
\textit{Trajectum}'s parameters are determined by Bayesian analysis of a variety of systems.
For the present work, we employed the  parameter sets from~\cite{vanderSchee:2025hoe} for our Pb+Pb and O+O simulations at $\sqrt{s_\text{NN}} = \unit[5.36]{TeV}$. These use T$_\text{R}$ENTo~\cite{Moreland:2014oya,Moreland:2018gsh} for the initial conditions and \textsc{Smash}~\cite{SMASH:2016zqf,Sjostrand:2007gs} for the hadronic transport. When running \textsc{Jewel} with Trajectum events as background the default assumption that the temperature rises linearly up to the initialisation time of the hydrodynamics is implemented also here.

\section{Set-up}

For the results in this paper we have used two main set-ups. One is \textsc{Jewel} combined with TGlauberMC as outlined above, which we run with different assumptions about the temperature profile at early times (i.e. before the initialisation time). In the default model there is a linear increase in the temperature from zero until $\tau_i$, where the expansion begins. The value of $T_i$ for this case was obtained by scaling the energy density obtained from T$_\text{R}$ENTo~\cite{Moreland:2014oya} for OO events at $0-10\%$ centrality. However, for this work, we have also explored two different scenarios: one in which the temperature is zero before $\tau_i$, and the other in which the initial temperature is constant and equal to its value at $\tau_i$ until the expansion begins. Since the goal is to understand the effect of coherence, a fourth scenario was considered: the temperature for the coherent case was kept the same as in the default scenario, whereas that for the incoherent evolution was retuned so as to give the same $R_{AA}$ as the coherent. The values of $T_i$ used in the different cases are summarized in Table~\ref{tab:tivalues}.

\begin{table}
    \begin{tabular}{|l|c|}
    \hline
    temperature profile & $T_i = T(\tau_i)\ [\mathrm{MeV}]$ \\
    \hline
    $T(\tau < \tau_i) = 0$ & 400 \\
    $T(\tau < \tau_i) = T_i$ & 280 \\
    $T(\tau < \tau_i) = T_i\tau/\tau_i$ & 320 \\
    $T(\tau < \tau_i) = T_i\tau/\tau_i$ retuned & 260 \\
    \hline
    \end{tabular}
    \caption{Values for the initial temperature $T(\tau_i)$ used for the different temperature profiles. The initial time is $\tau_i =\unit[0.6]{fm/c}$ for all profiles.}
    \label{tab:tivalues}
\end{table}

For the second set-up we generated medium profiles for central O+O collisions as well as peripheral Pb+Pb, both at $\sqrt{s_\text{NN}} = \unit[5.36]{TeV}$, using Trajectum with the parameters from~\cite{vanderSchee:2025hoe}. 

With both set-ups, we have generated di-jet events and analysed them using Rivet~\cite{Bierlich:2019rhm}, HepMC~\cite{Buckley:2019xhk}, FastJet~\cite{Cacciari:2011ma} and \textsc{Yoda}. The final state hadrons with $p_\perp > \unit[100]{MeV}$ falling within a rapidity interval of $|\eta|<3$ were clustered into jets of radius $R=0.4$ using the  anti-$k_\perp$ algorithm~\cite{Cacciari:2008gp}. To obtain the $R_{AA}$ for hadrons and jets the respective $p_\perp$ spectra were plotted along with that for a corresponding p+p run. As for the results of $v_2$, those were calculated as $v_2 = \langle\langle \cos(2[\phi_i-\Psi_2]) \rangle_i\rangle_{\text{event}}$, where $\phi_i$ are the azimuthal angle of the hadrons in the event and $\Psi_2^{\text{event}}$ is the corresponding participant plane angle for a given event. We compute $\Psi_2$ for a given event as:

\begin{equation}
    \Psi_2=\frac{1}{2}\arctan \left(\frac{\langle \sin(2\phi_\text{N}) \rangle}{\langle \cos(2\phi_\text{N}) \rangle}\right) + \frac{\pi}{2},
\end{equation}

where $\phi_\text{N}$ is the azimuthal angle of a wounded nucleon, and the averages are over the positions of all the wounded nucleons in the event.

\section{Results}
\label{sec:results}

\subsection{Sensitivity to temperature profile at early times}

\begin{figure}
    \includegraphics[width=\linewidth]{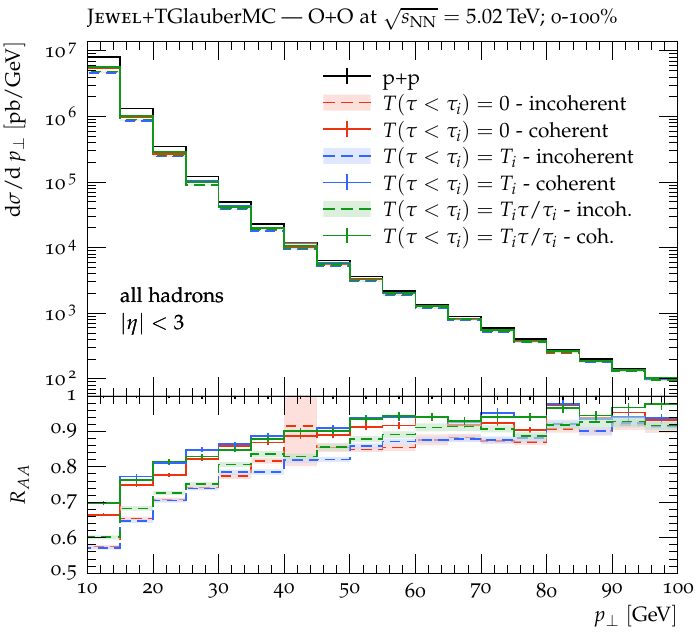}
    \caption{Hadron spectra and nuclear modification factors for the different temperature profiles with and without colour coherence. The initial temperatures for the different profiles are given in Table~\ref{tab:tivalues}.}
    \label{fig:glauber:hRAAprofiles}
\end{figure}

\begin{figure}
    \includegraphics[width=\linewidth]{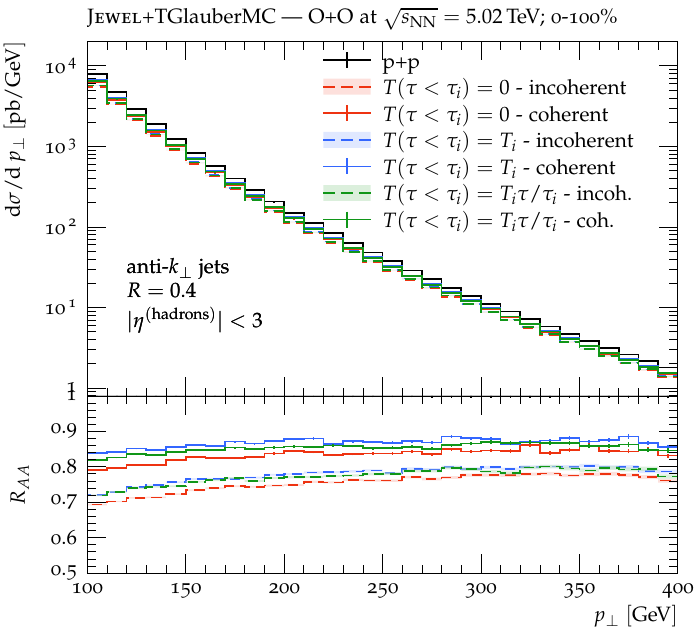}
    \caption{Jet spectra and nuclear modification factors for the different temperature profiles with and without colour coherence. The initial temperatures for the different profiles are given in Table~\ref{tab:tivalues}.}
    \label{fig:glauber:jRAAprofiles}
\end{figure}

In this section we present the results obtained with the set-up described above using \textsc{Jewel} with the initial temperature profiles based on TGlauberMC. \Fig{fig:glauber:hRAAprofiles} shows the $p_\perp$ spectra of hadrons obtained for events with and without coherence effects in the three different scenarios described above for the temperature before the initial time, and \Fig{fig:glauber:jRAAprofiles} shows the respective jet $p_\perp$ spectra. Since the goal here is to investigate how the colour coherence effects depend on the temperature profile at early times, the initial temperatures in the scenarios with zero or constant temperature before the initial time were chosen such that the nuclear modification factors are the same for the simulations without colour coherence. The naive expectation is that colour coherence is more important in scenarios with a higher temperature at early times. This trend is indeed observed in Figures~\ref{fig:glauber:hRAAprofiles} and \ref{fig:glauber:jRAAprofiles}, but the difference is not large. For jets, the nuclear modification factors with colour coherence are roughly 10\% higher than in the incoherent case. For the 0-10\% centrality class (not shown here) the difference is about 15\%, which is still somewhat smaller than the roughly 20\% observed in~\cite{ccpaper} for central Pb+Pb collisions.

\begin{figure}
    \includegraphics[width=\linewidth]{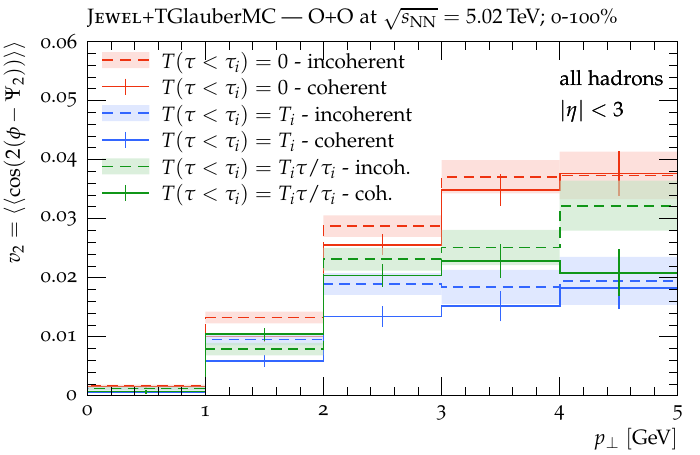}
    \caption{Hadron $v_2$ for the different temperature profiles with and without colour coherence in bins of hadron transverse momentum. $v_2$ is calculated as $v_2 = \langle \langle \cos(2(\phi - \Psi_2)) \rangle \rangle$, where $\Psi_2$ is the participant plane angle obtained from the Glauber model. The averages are taken first over the hadrons in a given event, and then over the events.}
    \label{fig:glauber:hv2profiles}
\end{figure}

\begin{table}
    \begin{tabular}{|l|p{1.5cm}|c|}
    \hline
    temperature profile & colour coherence & jet $v_2$ \\
    \hline
    $T(\tau < \tau_i) = 0$ & incoh. & $0.024 \pm 0.001$\\
                           & coh. & $0.018 \pm 0.001$\\
    $T(\tau < \tau_i) = T_i$ & incoh. & $0.012 \pm 0.001$\\
                             & coh.& $0.008 \pm 0.001$ \\
    $T(\tau < \tau_i) = T_i\tau/\tau_i$ & incoh. & $0.015 \pm 0.001$\\
                                        & coh. & $0.012 \pm 0.001$\\
    $T(\tau < \tau_i) = T_i\tau/\tau_i$ retuned & incoh. & $0.012 \pm 0.001$ \\
    \hline
    \end{tabular}
    \caption{Jet $v_2$ for the linear temperature profiles with and without colour coherence for jets with $p_\perp > \unit[60]{GeV}$. The initial temperature of the incoherent simulation was adjusted to yield the same nuclear modification factor as the coherent one. $v_2$ is calculated as $v_2 = \langle \langle \cos(2(\phi - \Psi_2)) \rangle \rangle$, where $\Psi_2$ is the participant plane angle obtained from the Glauber model. The averages are taken first over the hadrons in a given event, and then over the events.}
    \label{tab:jetv2s}
\end{table}

\begin{figure}
    \includegraphics[width=\linewidth]{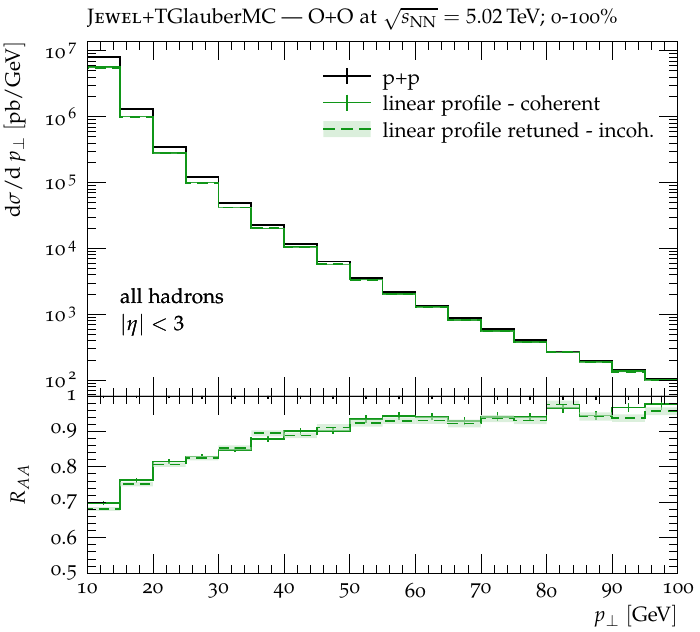}
    \caption{Hadron spectra and nuclear modification factors for the linear temperature profiles with and without colour coherence, where the initial temperature of the incoherent simulation was adjusted to yield the same nuclear modification factor as the coherent one.}
    \label{fig:glauber:hRAAv2comp}
\end{figure}

\begin{figure}
    \includegraphics[width=\linewidth]{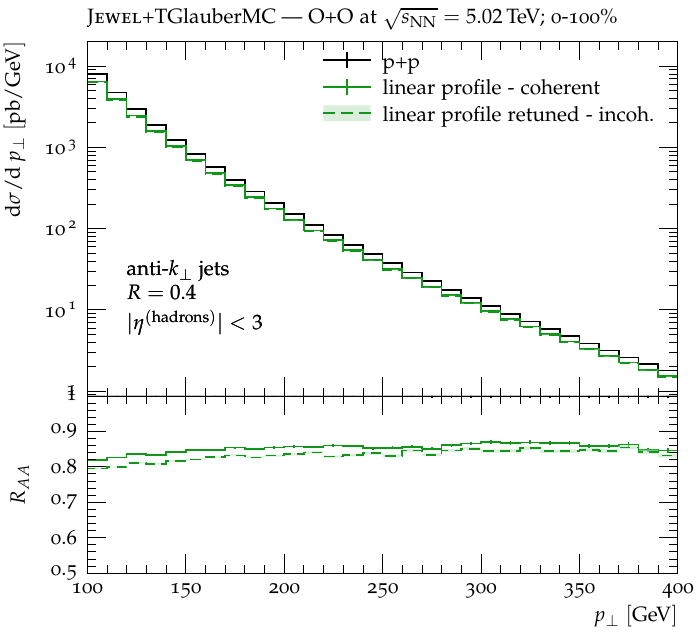}
    \caption{Jet spectra and nuclear modification factors for the linear temperature profiles with and without colour coherence, where the initial temperature of the incoherent simulation was adjusted to yield the same nuclear modification factor as the coherent one.}
    \label{fig:glauber:jRAAv2comp}
\end{figure}

\begin{figure}
    \includegraphics[width=\linewidth]{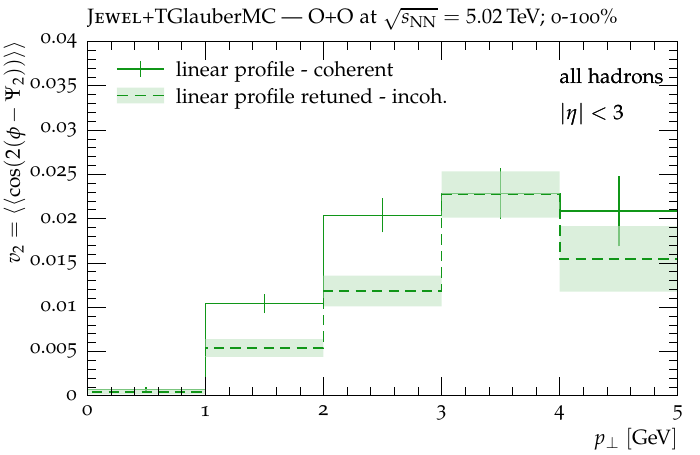}
    \caption{Hadron $v_2$ for the linear temperature profiles with and without colour coherence in bins of hadron transverse momentum. The initial temperature of the incoherent simulation was adjusted to yield the same nuclear modification factor as the coherent one. $v_2$ is calculated as $v_2 = \langle \langle \cos(2(\phi - \Psi_2)) \rangle \rangle$, where $\Psi_2$ is the participant plane angle obtained from the Glauber model. The averages are taken first over the hadrons in a given event, and then over the events.}
    \label{fig:glauber:hv2v2comp}
\end{figure}

At high transverse momentum an azimuthal anisotropy is induced in non-central collisions by the path-length dependence of energy loss. Figure~\ref{fig:glauber:hv2profiles} and Table~\ref{tab:jetv2s} show the elliptic flow coefficient $v_2$ for the different temperature profiles for hadron and jets, respectively. We would like to stress that the hadrons entering the calculation of $v_2$ here are semi-soft fragments of hard scattering processes (the $\hat p_{\perp,\text{min}}$ on the hard scattering matrix element is \unit[6]{GeV}), and not inclusive hadrons at a given $p_\perp$. The anisotropy is largest for the profile with zero temperature up to $\tau_i$ and smallest for the one with $T(\tau < \tau_i) = T_i$. This is expected, since the system as seen by hard partons at very early times is almost isotropic and the anisotropy is built up mostly at later times. This has also been observed in~\cite{Andres:2016iys}. There is also an ordering (more significant for jets than for hadrons) between coherent and incoherent scenarios with the coherent ones having a slightly smaller $v_2$. 
However, a direct comparison is hindered by the fact that the $R_\text{AA}$ is different in the coherent and incoherent cases. Therefore, we made a simulation with the default temperature profile (with a linear increase at early times), where the initial temperature was lowered in the incoherent case to yield the same nuclear modification factor as in the coherent case. The corresponding nuclear modification factors are shown in Figures~\ref{fig:glauber:hRAAv2comp} and \ref{fig:glauber:jRAAv2comp}, and the elliptic flow coefficients in Figure~\ref{fig:glauber:hv2v2comp} and Table~\ref{tab:jetv2s}. For jets, $v_2$ is the same in both bases, but for hadrons $v_2$ is somewhat larger in the coherent case. This can be understood by noting that colour coherence suppresses scatterings at early times. Therefore the suppression is built up at later times compared to the incoherent scenario, which leads to a slightly larger $v_2$.  

\subsection{Sensitivity to system shape}

\begin{figure}
    \includegraphics[width=\linewidth]{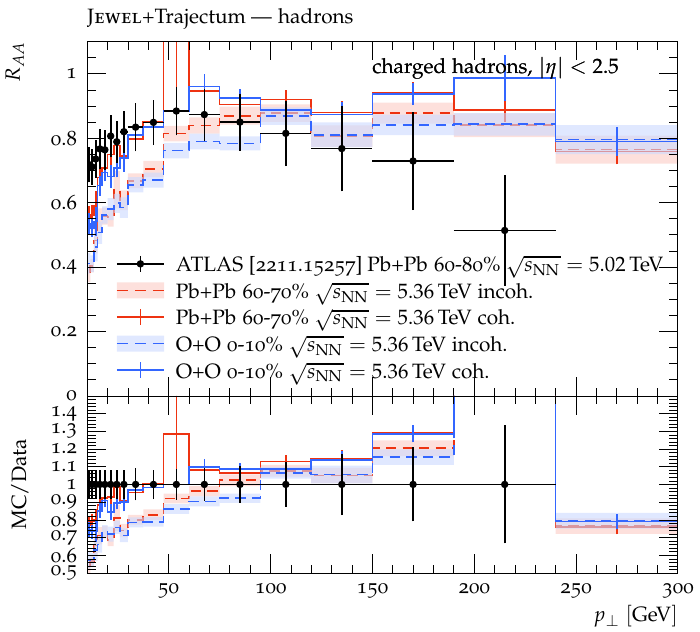}
    \caption{Nuclear modification factor for charged hadrons with $|\eta| < 2.5$. The \textsc{Atlas} data~\cite{ATLAS:2022kqu} are for Pb+Pb collisions at $\sqrt{s_\text{NN}} = \unit[5.02]{TeV}$ with 60-80\% centrality. \textsc{Jewel}+Trajectum results are shown for Pb+Pb collisions with 60-70\% centrality and O+O collisions with 0-10\% centrality, both at $\sqrt{s_\text{NN}} = \unit[5.36]{TeV}$.}
    \label{fig:hydro:hRAA}
\end{figure}

\begin{figure}
    \includegraphics[width=\linewidth]{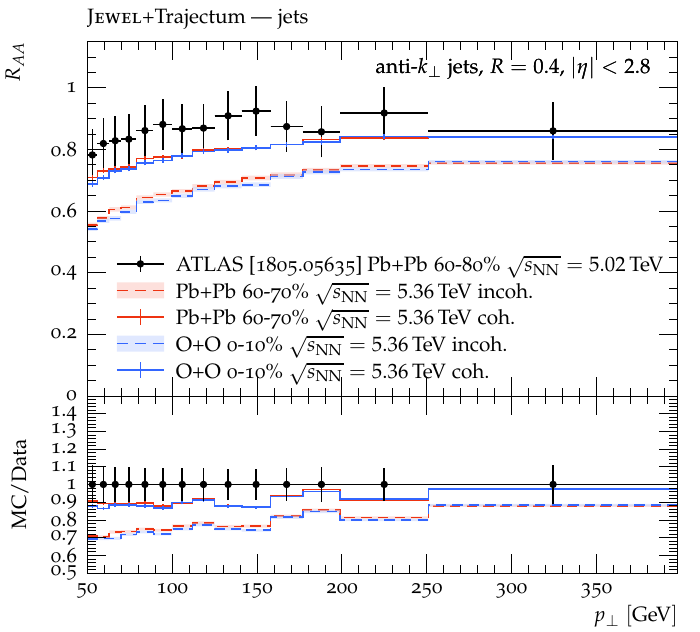}
    \caption{Nuclear modification factor for anti-$k_\perp$ jets, $R=0.4$ with $|\eta| < 2.8$. The \textsc{Atlas} data~\cite{ATLAS:2018gwx} are for Pb+Pb collisions at $\sqrt{s_\text{NN}} = \unit[5.02]{TeV}$ with 60-70\% centrality. \textsc{Jewel}+Trajectum results are shown for Pb+Pb collisions with 60-70\% centrality and O+O collisions with 0-10\% centrality, both at $\sqrt{s_\text{NN}} = \unit[5.36]{TeV}$.}
    \label{fig:hydro:jRAA}
\end{figure}

This section aims at comparing systems of comparable size but different shapes. This requires a better medium model, because we chose to use the multiplicity as measure for the size, and the simplistic medium model used in the previous section cannot provide that. Therefore, we here use temperature and velocity profiles generated by \textit{Trajectum} as medium model for \textsc{Jewel}. We compare 0-10\% central O+O collisions (multiplicity of charged particles with $p_\perp > \unit[200]{MeV}$ in $\vert\eta\vert < 0.9$ is 175) to 60-70\% central Pb+Pb collisions (charged particle multiplicity 178) at $\sqrt{s_\text{NN}}=\unit[6.36]{TeV}$. These two systems have very different eccentricities and transverse flow patterns, but similar hydrodynamic initialisation times (\unit[0.355]{fm/c} for Pb+Pb and \unit[0.352]{fm/c} for O+O). Since colour coherence suppresses early scatterings one could speculate whether a difference between these two systems becomes visible when colour coherence is turned on. However, this is not the case for the hadron and jet $R_\text{AA}$ shown in Figures~\ref{fig:hydro:hRAA} and \ref{fig:hydro:jRAA}, respectively. We would like to stress that a quantitative comparison to the \textsc{Atlas} data is not possible due to the different beam energy and, in the case of the hadron $R_\text{AA}$, centrality selections.  Qualitatively, the result from \textsc{Jewel}+\textit{Trajectum} for peripheral Pb+Pb collisions with colour coherence are in reasonable agreement with the measurement, while the incoherent scenario shows a too strong suppression. This finding is in line with what is observed in~\cite{ccpaper} for central Pb+Pb collisions.

\section{Conclusions}

We have studied the effect of color coherence on jet and hadron $R_{AA}$ and $v_2$ in O+O and peripheral Pb+Pb collisions. We found that the role of coherence is more important when the temperature at early times is higher. This is expected since the suppression of re-scattering due to colour coherence is strongest at early times, when the colour dipoles are still small in transverse size. It was found that $v_2$ shows the expected correlation with $R_\text{AA}$, where a stronger suppression leads to a larger $v_2$. Therefore, in order to isolate the effect of colour coherence on $v_2$, we construct a setup where $R_\text{AA}$ is very similar with and without colour coherence. We found that, in this case, $v_2$ tends to be larger with coherence effects turned on. This happens because scatterings on average take place later when colour coherence is turned on, when the system is more anisotropic. Colour coherence is thus a mechanism that enhances $v_2$ relative to $R_\text{AA}$ and could therefore be part of the answer to the question why a sizable $v_2$ is observed in small systems while the nuclear modification factors do not show a significant suppression. 

To further elucidate the role of spatial anisotropy and anisotropic flow of the background medium we compare central O+O to peripheral Pb+Pb collisions at the same charged particle multiplicity. It turned out that in spite of the very different shapes of these two systems the nuclear modification factors for hadrons and jets are the same. This suggests that any non-linearities are small enough that the differences average out in $R_\text{AA}$.

\section*{Acknowledgments}

The authors would like to thank Govert Nijs for help with getting set up with \textit{Trajectum} and parameter choices.
This study is part of a project that has received funding from the European Research Council (ERC) under the European Union's Horizon 2020 research and innovation programme  (Grant agreement No. 803183, collectiveQCD). 
This work is also supported by the DST/NRF in South Africa under Thuthuka grant number TTK240313208902.

\bibliographystyle{unsrtnat}
\bibliography{refs}

\end{document}